\documentclass[preprint,pre,floats,aps,amsmath,amssymb]{revtex4-2} 

\usepackage{color,bm}
\usepackage{graphicx}

\newcommand{\lr}[1]{\left(#1\right)}
\newcommand{\mean}[1]{\left\langle#1\right\rangle}
\newcommand{\vare}{\varepsilon}

\begin{document}

\title{Non-equilibrium fluctuations of the direct cascade in Surface Quasi Geostrophic turbulence}

\author{V.J. Valad\~{a}o$^{1,2}${\footnote{Corresponding author: victor.dejesusvaladao@unito.it}},
  T. Ceccotti$^{1}$, G. Boffetta$^{1,2}$, and S. Musacchio$^{1,2}$}
\affiliation{
  $^{1}$Dipartimento di Fisica - Università degli Studi di Torino, Via P. Giuria, 1, 10125 Torino, Italy.\\
  $^{2}$INFN Sezione di Torino, Via P. Giuria 1, 10125 Torino, Italy.}
\vspace{1.5cm}

\begin{abstract}

We study the temporal fluctuations of the flux of surface potential energy
in Surface Quasi-Geostrophic (SQG) turbulence.
By means of high-resolution, direct numerical simulations of the SQG
model in the regime of forced and dissipated cascade of temperature variance,
we show that the instantaneous imbalance in the energy budget originate a
subleading correction to the spectrum of the turbulent cascade.  Using a
multiple scale approach combined with a dimensional closure we derive a
theoretical prediction for the power-law behavior of the corrections, which
holds for a class of turbulent transport equations known as
$\alpha$-turbulence. Further, we develop and apply a method to disentangle 
the equilibrium and non-equilibrium contribution in the instantaneous 
spectra, which can be generalized to other turbulent systems.

\end{abstract}

 \maketitle

\section{Introduction}
\label{introduction}

The Surface Quasi-Geostrophic (SQG) equation has been proposed as a model
to describe the flow determined by the conservation of buoyancy at the surface
of a stratified fluid in rotation \cite{blumen1978uniform,salmon1998lectures}.
Within the framework of quasi-geostrophic flows, 
the SQG has been used as a model for the dynamics of the Earth's atmosphere at
the tropopause~\cite{juckes1994quasigeostrophic},
for the surface dynamics of the oceans~\cite{lapeyre2006dynamics},
and, more recently, for the convective motions in Jupiter's
atmosphere~\cite{siegelman2022moist}.

Besides its interest in geophysical and astrophysical applications,
the SQG equation is appealing also for theoretical studies of turbulence.
Formally, the SQG can be seen as a specific instance of a broader class of 
two-dimensional (2D) models, the so-called $\alpha$-turbulence 
models~\cite{pierrehumbert1994spectra},
which generalizes the 2D Navier-Stokes (NS) equation and 
describe the transport of an active scalar field by a 2D 
incompressible flow.
The latter is determined by a functional relation between the stream function and the scalar field itself.
In analogy with the case of the 2D NS equation,
the $\alpha$-turbulence possesses two quadratic inviscid invariants,
which gives rise to a double cascade phenomenology, with an inverse cascade of one invariant toward large scales and a direct cascade of the other toward small scales. 

In the SQG case, the transported scalar field corresponds to the potential surface temperature,
and the two invariants are the total energy and the surface potential energy. 
A peculiarity of the SQG model is that the hypothesis of a constant flux
of the surface potential energy toward small scales 
leads to the dimensional prediction for a Kolmogorov-like spectrum
$k^{-5/3}$~\cite{blumen1978uniform,pierrehumbert1994spectra}
in the range of wavenumbers corresponding to the direct cascade.
Therefore, the SQG model displays distinctive features of both 2D
and 3D turbulent flows.
For this reason, it attracted the attention of the scientific community
interested in the statistical properties of turbulent cascades and transport in
turbulent
flows~\cite{held1995surface,celani2004active,lapeyre2017surface,foussard2017relative}
as well as in the development of singularities~\cite{constantin1994formation,constantin1999behavior}
and spontaneous stochasticity~\cite{valade2024anomalous}. 

It is worth to notice that, in general, the $k^{-5/3}$ prediction 
holds only for the time-averaged energy spectrum,
since it relies on the assumption of statistical stationarity of the system. 
In a turbulent flow, instantaneous imbalance can occur
between the injection at large-scale due to the external forcing and the small-scale dissipation.
In the case of 3D NS turbulence, theoretical studies
performed with a two-scale direct-interaction approximation method~\cite{yoshizawa1994nonequilibrium} 
and with the multiple-scale perturbation method~\cite{woodruff2006multiple}
showed that the temporal fluctuations of the energy flux results in a sub-leading correction for the slope
of the instantaneous energy 
spectra~\cite{yoshizawa1994nonequilibrium,woodruff2006multiple,berti2023mean}.

In this paper, we address the issue of the non-equilibrium correction to the
energy spectrum of the direct cascade of surface potential energy in the SQG
model. We derive a general prediction for the correction to the spectral 
slope of the direct cascades of the $\alpha$-turbulence model which depends 
on the fluctuations of the small-scale dissipation rate.  By means of numerical
simulations at high resolution, we verify the prediction in the SQG case
and we discuss the role of temporal fluctuations in the statistics of
the turbulent flow. An exact equation for the flux of the transported field,
a generalization of the Karman-Howarth-Monin equation of turbulence,
is derived in the Appendix.

\section{Surface Quasi-Geostrophic model}
\label{sec2}
  
The governing equation of the SQG model
is written in terms of the surface temperature field $\theta({\bm x},t)$
as~\cite{pierrehumbert1994spectra}
\begin{equation}
\label{eq1}
\partial_t\theta+{\bm v} \cdot {\bm \nabla}\theta = \kappa\Delta\theta + f\;,
\end{equation}
where $\kappa$ is the diffusivity and $f$ represents a large-scale forcing.
The two-dimensional, incompressible velocity field ${\bm v}({\bm x},t) = (-\partial_y \psi,\partial_x \psi)$
is related to the scalar field $\theta$ via the stream function by 
$\psi = |\Delta|^{-1/2}\theta$.
In the $\alpha$-turbulence model, the relation between the stream function 
and the scalar field is generalized as 
$\psi = |\Delta|^{-\alpha/2}\theta$~\cite{pierrehumbert1994spectra}.
Clearly, the SQG model corresponds to the case $\alpha=1$, while for 
$\alpha=2$ one recovers the equation for the scalar vorticity of 2D 
NS equation.  

In Fourier space, the relation between the velocity $\hat{\bm v}_{\bm k}$
and scalar field $\hat{\theta}_{\bm k}$ can be expressed, in the general case,
as 
\begin{equation}
\label{eq2}  
\hat{\bm v}_{\bm k} = \left(-{i k_y \over k^\alpha}, {i k_x \over k^\alpha}\right) \hat{\theta}_{\bm k}
\end{equation}
from which we note that in the SQG case the fields $\theta$ and 
${\bm v}$ have the same dimension. 

In the absence of the forcing and dissipation ($f=0$, $\kappa =0$)
the SQG model has two conserved quantities~\cite{blumen1978uniform},
the total vertically integrated energy (VIE) 
\begin{equation}
\label{eq3}
E=\frac{1}{2} \langle \psi \theta \rangle
\end{equation}
and the surface potential energy (SPE)
\begin{equation}
\label{eq4}
P= \frac{1}{2} \langle \theta^2 \rangle
\end{equation}
where the brackets $\langle \cdot \rangle$ stand for the spatial average.
Alternatively, the conserved quantities $E$ and $P$
are also referred to as generalized energy and enstrophy
\cite{pierrehumbert1994spectra} 
because of their resemblance to the inviscid invariants of 2D Navier-Stokes (NS) turbulence.
As a consequence of the relation between the fields $\theta$ and ${\bm v}$,
the SPE is equivalent to the surface kinetic energy (SKE) since
\begin{equation}
\label{eq5}
\langle \theta^2 \rangle = \sum_{\bm k} |\hat\theta_k|^2 = \langle |{\bm v}|^2 \rangle
\end{equation}

When the SQG flow is sustained by an external forging $f$,
with a characteristic forcing scale $\ell_f$, 
it develops two turbulent 
cascades~\cite{blumen1978uniform,pierrehumbert1994spectra}.
The SPE is transferred mostly toward small-scales $\ell < \ell_f$ 
giving rise to a direct cascade of variance of temperature fluctuations,
while the VIE is transferred toward large-scales $\ell > \ell_f$ 
by an inverse energy cascade.
The two cascades are stopped by dissipation mechanisms acting at small-scales
(e.g. diffusivity or viscosity) and large scale (such as friction).

In this work, we will focus on the range of scales
comprised between the forcing scale $\ell_f$ and the 
diffusive scale $\ell_\kappa$, 
corresponding to the direct cascade of SPE.
The balance of surface potential energy is
$dP/dt = \varepsilon_I - \varepsilon_\kappa$,
where $\varepsilon_I = \langle \theta f \rangle$ is the SPE input rate
and $\varepsilon_\kappa = \kappa \langle |{\bm \nabla} \theta|^2 \rangle$
is the SPE dissipation rate at small scales. 
The assumption of stationarity implies that, on average,
$\varepsilon_I = \varepsilon_\kappa$,
and that the input and dissipation rates are equal to the flux 
$\varepsilon$ of SPE in the cascade. 

The further assumption of statistical homogeneity and isotropy allows
to derive an exact relation for a mixed structure function
which involves both the scalar and velocity increments in the range of scales
of the direct cascade. 
Let us define the increments of the scalar field
$\delta \theta(\ell) = \theta({\bm x}+{\bm \ell},t)-\theta({\bm x},t)$
and the increments of the $i$-component of the velocity field
$\delta v_i(\ell) = v_i({\bm x}+{\bm \ell},t)-v_i({\bm x},t)$.
In the range of scales $\ell_{\kappa} < \ell < \ell_f$, the 
mixed longitudinal structure function
$S^L(\ell) = \langle (\delta \theta)^2 \delta v_i) \rangle\ell_i/\ell$
satisfies the following relation (see Appendix\ref{app:a}):
\begin{equation}
S^L(\ell) = -2 \varepsilon \ell\;.
\label{eq6}
\end{equation}
Note that this relation is valid not only for the SQG case,
but for all the direct cascades of the $\alpha$-turbulence model 
with $\alpha > 0$.

Under the hypothesis of scale invariance of the system,
the statistics of the scalar increments depends on the scale as 
$\delta \theta (\ell) \sim \ell^h$ where the scaling exponent 
$h$ is determined as follows.
From the relation \eqref{eq2} between $\theta$ and ${\bm v}$, 
the scaling of the velocity increments is 
$\delta v(\ell) \sim \ell^{\alpha-1}\delta\theta(\ell) \sim\ell^{h+\alpha-1}$.
Inserting these scaling relations in (\ref{eq6}) one obtains
\begin{equation}
\label{eq7}  
h=\frac{2-\alpha}{3}
\end{equation}
which leads to the prediction for the spectrum of the direct cascade \cite{pierrehumbert1994spectra}
\begin{equation}
\label{eq8}
P(k) \sim \varepsilon^{2/3}k^{\frac{2\alpha-7}{3}}\ .\
\end{equation}
In the SQG case ($\alpha=1$) 
the scaling relations for the temperature and velocity fluctuations
are the same
$\delta \theta(\ell) \sim \delta v(\ell) \sim \varepsilon^{1/3} \ell^{1/3}$
and the prediction for the spectrum $P(k)\sim \varepsilon^{2/3} k^{-5/3}$ 
of temperature variance is formally identical to that of kinetic energy
in classical Kolmogorov 3D turbulence~\cite{frisch1995turbulence}. Following the same rationale as in 3D turbulence, one can define a diffusive scale $\ell_\kappa = (\kappa^3/\vare)^{1/4}$ representing the smallest active scale in the problem such as the Kolmogorov scale. 
  
\section{Non-equilibrium spectral corrections}
\label{sec3}

The dimensional arguments discussed in the previous section
require the assumption of statistical stationarity of the system.
In particular, it is necessary to assume that the energy input, the flux of the cascade
and the small-scale dissipation are equal so that the system is at equilibrium.
In a turbulent flow, this balance is realized only on average and
the instantaneous imbalance between injection and dissipation occurs
because of the temporal fluctuations of the forcing and the intermittent 
nature of the small-scale dissipation. 
As a consequence, the prediction (\ref{eq8}) is valid only for time-averaged spectra. 
The effects of the non-equilibrium, temporal fluctuations of the flux
in the direct cascade of $\alpha$-turbulence can be
investigated using simple heuristic arguments in analogy
with the approach adopted in 3D turbulence~\cite{bos2017dissipation}.

The time evolution of the spectrum $P(k,t)$ related to the invariant that cascades to small-scales is governed by \cite{batchelor1953theory} 
\begin{equation}
\label{eq9} 
\partial_t P(k,t) = F(k,t) - \partial_k \Pi(k,t) - D(k,t) \, , 
\end{equation}
where $\Pi(k,t)$ is the flux of the turbulent cascade,   
$F(k,t)$ is the production spectrum due to the external force and 
$D(k,t) = 2 \kappa k^2 P(k,t)$ is the dissipation spectrum.
For the flux term, we adopt a simple dimensional closure,
generalization of that used for 3D turbulence \cite{kovasznay1948spectrum}.
From the dimensional relation $\Pi(k,t) = k P(k,t)/\tau_k$,
using $\tau_k = k^{\alpha-5/2} P(k,t)^{-1/2}$ for the eddy turnover time
at the scale $\ell = 1/k$, one gets
\begin{equation}
\Pi(k,t) = C k^{\frac{7-2\alpha}{2}} P(k,t)^{3/2} \, .
\label{eq10} 
\end{equation}

In order to consider the effect of temporal fluctuations out of equilibrium, 
let us expand the spectrum and the flux 
as $P(k,t) = P_0(k,t) + \epsilon P_1(k,t)$ and
$\Pi(k,t) = \Pi_0(k,t) + \epsilon \Pi_1(k,t)$ 
where the first terms represent the instantaneous equilibrium values
while the seconds are the first-order non-equilibrium correction.
The small parameter $\epsilon \ll 1$ reflects the assumption that the 
correction is smaller than the equilibrium solution.
We also assume that temporal fluctuations are on a slow time scale 
of the same order of the corrections and therefore 
we replace the time derivative in (\ref{eq9}) with $\epsilon \partial_t$.

Inserting the above expansion in (\ref{eq9}), neglecting the production and
dissipation terms in the inertial range, we get at the leading
order $\epsilon^0$, $0 = -\partial_k \Pi_0(k,t)$ 
which implies that the equilibrium flux $\Pi_0$ is independent on $k$,
i.e. $\Pi_0(k,t)=\varepsilon(t)$. Using now the closure (\ref{eq10}) we 
obtain for the equilibrium spectrum again the prediction (\ref{eq8})
\begin{equation}
P_0(k,t) = C_0 \varepsilon(t)^{2/3} k^{\frac{2\alpha-7}{3}} \,,
\label{eq11}
\end{equation}
with $C_0 = C^{-2/3}$.

At the first order in $\epsilon$, (\ref{eq9}) gives 
$\partial_t P_0(k,t) = - \partial_k \Pi_1(k,t)$
and assuming a power-law form for the correction $P_1(k,t) = g(t) k^{\beta}$,
we finally obtain the prediction for the first-order correction
\begin{equation} 
P_1(k,t) = C_1 \varepsilon(t)^{-2/3} \dot{\varepsilon}(t)
k^{\frac{4\alpha-11}{3}} \, ,
\label{eq12}
\end{equation} 
where $C_1 = 2 C_0^2/3(2-\alpha)$.
Note that for $0 < \alpha < 2$ the spectral slope of $P_1$ is steeper than 
that of $P_0$, therefore, the correction is subdominant at large wavenumbers.
In particular, for the SQG case ($\alpha=1$) we obtain the prediction that 
the non-equilibrium correction has a spectral exponent $-7/3$. 
Conversely, in the case of the direct cascade of enstrophy in 2D NS 
turbulence ($\alpha=2$) the coefficient $C_1$ diverges, 
spectral exponents for the equilibrium and non-equilibrium spectra are 
the same, and the perturbative expansion is inconsistent.

From (\ref{eq11}-\ref{eq12}) we can write the ratio between the non-equilibrium and equilibrium spectra as
\begin{equation} 
\frac{P_1(k,t)}{P_0(k,t)} =
\frac{2 C_0}{3(2-\alpha)} \varepsilon(t)^{-4/3} \dot{\varepsilon}(t) k^{\frac{2}{3}(\alpha-2)} =
\frac{2 C_0}{3(2-\alpha)} \frac{\tau_k}{\tau_\varepsilon}\, ,
\label{eq13}
\end{equation}
which is the ratio between the eddy turnover time 
$\tau_k = \varepsilon(t)^{-1/3}k^{(2\alpha-4)/3}$
and the time scale of temporal fluctuations of the flux 
$\tau_\varepsilon=\varepsilon(t)/\dot{\varepsilon}(t)$ 
which has been assumed to be large when compared to $\tau_f$.
This relation justifies a posteriori the perturbative assumption
based on the single parameter $\epsilon$.

In the statistically stationary regime, a simple procedure allows to identify
the equilibrium and non-equilibrium components of the instantaneous spectra
$P(k,t)$.  
By taking the time average over time scales much longer than 
$\tau_\varepsilon$, we have  $\langle P_1(k,t) \rangle_t =0$ 
since (\ref{eq12}) can be written as a total time derivative.
Therefore we have 
$\langle P(k,t) \rangle_t = \langle P_0(k,t) \rangle_t$. 
Moreover, multiplying $P(k,t)$ by the $\varepsilon(t)^{-2/3}$
and computing the time average, we have
$\langle P(k,t) \varepsilon(t)^{-2/3} \rangle_t =
\langle P_0(k,t) \varepsilon(t)^{-2/3} \rangle_t +
\langle P_1(k,t) \varepsilon(t)^{-2/3} \rangle_t$.
The first term $P_0(k,t) \varepsilon(t)^{-2/3}=C_0 k^{(2 \alpha-7)/3}$ 
is time independent while, again, the second term vanishes since it is
the time average of a total time derivative. 
In conclusion, the leading order term can be computed as 
\begin{equation}
P_0(k,t) = \varepsilon(t)^{2/3} \langle P(k,t) \varepsilon(t)^{-2/3} \rangle_t
\label{eq14}
\end{equation}
and the subleading correction as the difference
\begin{equation}
P_1(k,t) = P(k,t) - \varepsilon(t)^{2/3} \langle P(k,t) \varepsilon(t)^{-2/3} \rangle_t \, .
\label{eq15}
\end{equation}

\section{Numerical simulations and Results}
\label{sec4}

In order to test the predictions derived in the previous section, we 
performed direct numerical simulations of the SQG equation 
\eqref{eq1} in a doubly-periodic square box of size $L=2\pi$,
discretized on a regular grid of $N^2=8192^2$ collocation points. 
The simulations are done with a fully-dealiased pseudospectral code
\cite{boyd2001chebyshev}
with fourth-order Runge-Kutta time scheme implemented on GPU with 
OpenACC directives \cite{valadao2024spectrum}.
The flow is sustained by a random, white-in-time forcing $f$,
which provides an average injection rate of SPE $\varepsilon_I=24$. 
To maximize the range of scale available in the direct cascade,
the forcing is active only on a narrow shell of wavenumbers 
$3 \le |{\bm k}_f| \le 4$,
which defines the forcing scale $\ell_f = 2\pi/k_f$. 
The inverse cascade is completely suppressed by means of an additional
(hypo)-friction term $-\mu\nabla^{-2}\theta$ on the right-hand side 
of \eqref{eq1}. The coefficients of the dissipative terms are $\kappa=10^{-4}$ 
and $\mu=1$ such that the simulation is resolved until $k_{max}\ell_\kappa\approx1.5$. 

The Reynolds number is a delicate quantity to be defined in the double cascade scenario since numerically, it will always incorporate the effects of the existence of an inverse cascade. To avoid such an influence, we define the diffusive Reynolds number in terms of forcing quantities as $Re_\kappa=\vare_I^{1/3}\ell_f^{4/3}\kappa^{-1}\approx 63000$. All the results are made dimensionless using the characteristic scale
$\ell_f$ and time $\tau_f=\varepsilon_I^{-1/3} \ell_f^{2/3}$ of the 
forcing.

\begin{figure}[h]
\centering\includegraphics[width=0.80\linewidth]{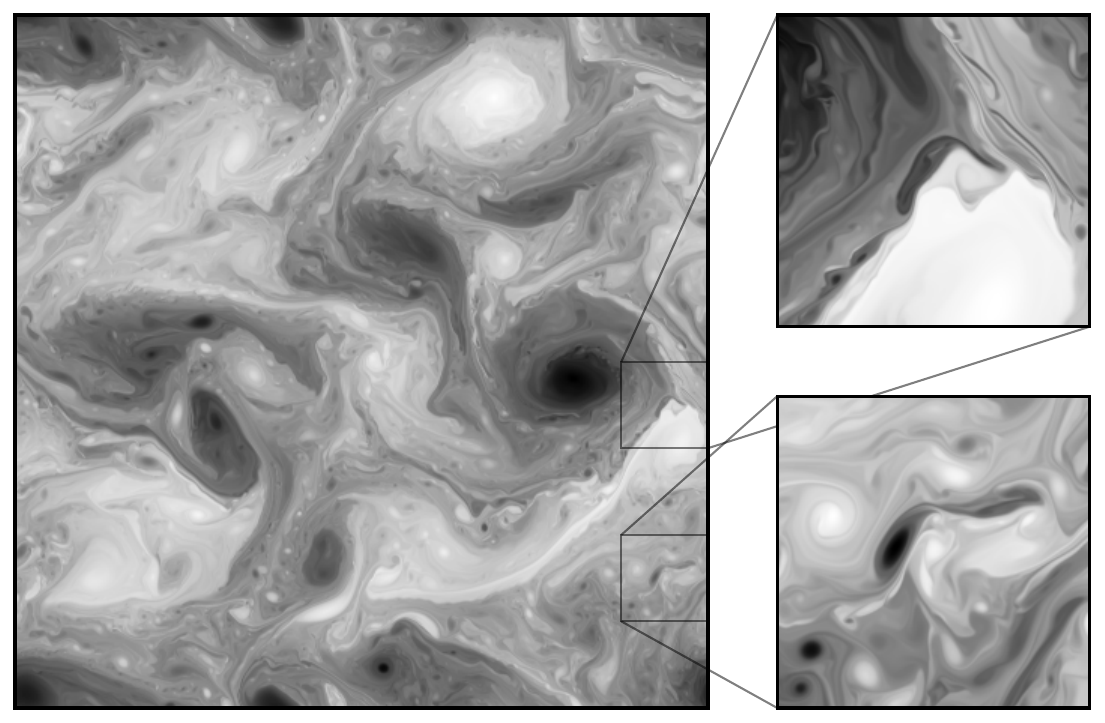}
\caption{Snapshot of a numerical solution for the scalar field $\theta$. Zoomed areas emphasize the presence of coherent vortices and rough filaments at small scales.}
\label{fig1}
\end{figure}

The simulations are initialized with a null scalar field $\theta({\bm x},t)$.
After an initial transient, the system eventually develops a statistically stationary turbulent regime
in which the typical aspect of the scalar field $\theta({\bm x},t)$
is shown in Fig.~\ref{fig1}.
From the figure, one notes that the statistics of SQG turbulence 
differs from the typical solution of a 2D NS equation since the 
instability of temperature filaments produces a rough velocity field
characterized by the formation of vortices at all scales.

\begin{figure}[h]
    \includegraphics[width=0.7\linewidth]{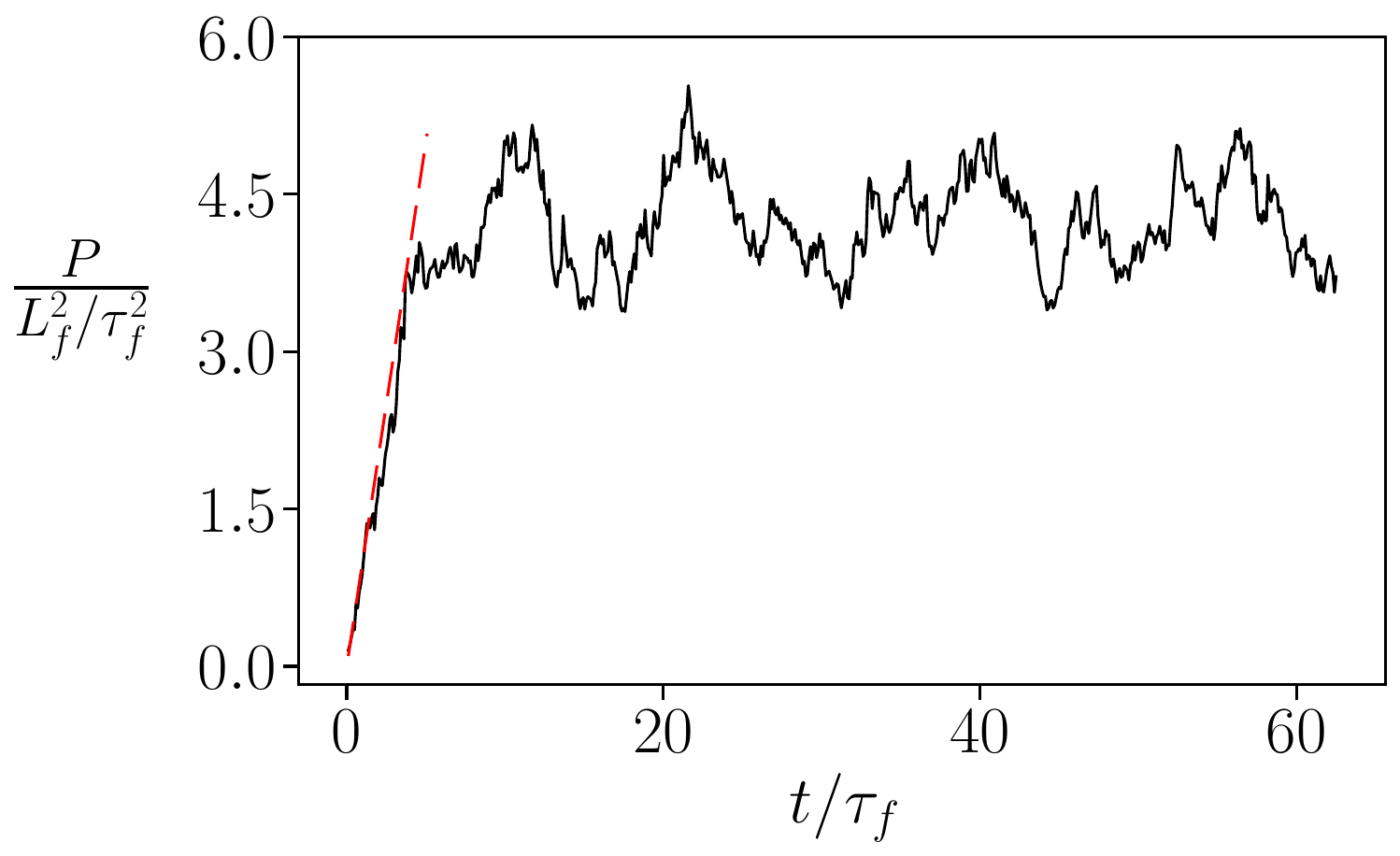}
\caption{Time evolution of the $P$. The dashed line shows the transient regime. \label{fig2}}
\end{figure}

The temporal evolution of temperature variance $P(t)$ is reported in Fig.~\ref{fig2}.
After the initial stage of the evolution ($t\lesssim 5\tau_f$),
in which $P(t)$ grows linearly in time with the input rate provided by forcing $P(t) = \varepsilon_It$,
the system reaches a statistically stationary regime.
It is worth emphasizing the existence of instantaneous large fluctuations,
with a typical amplitude of about $20\%$ of the mean value $\mean{P}_t$,
which indicates that even though the system is statistically stationary over long times, 
instantaneously it is always out of equilibrium.

\begin{figure}[h]
    \includegraphics[width=0.7\linewidth]{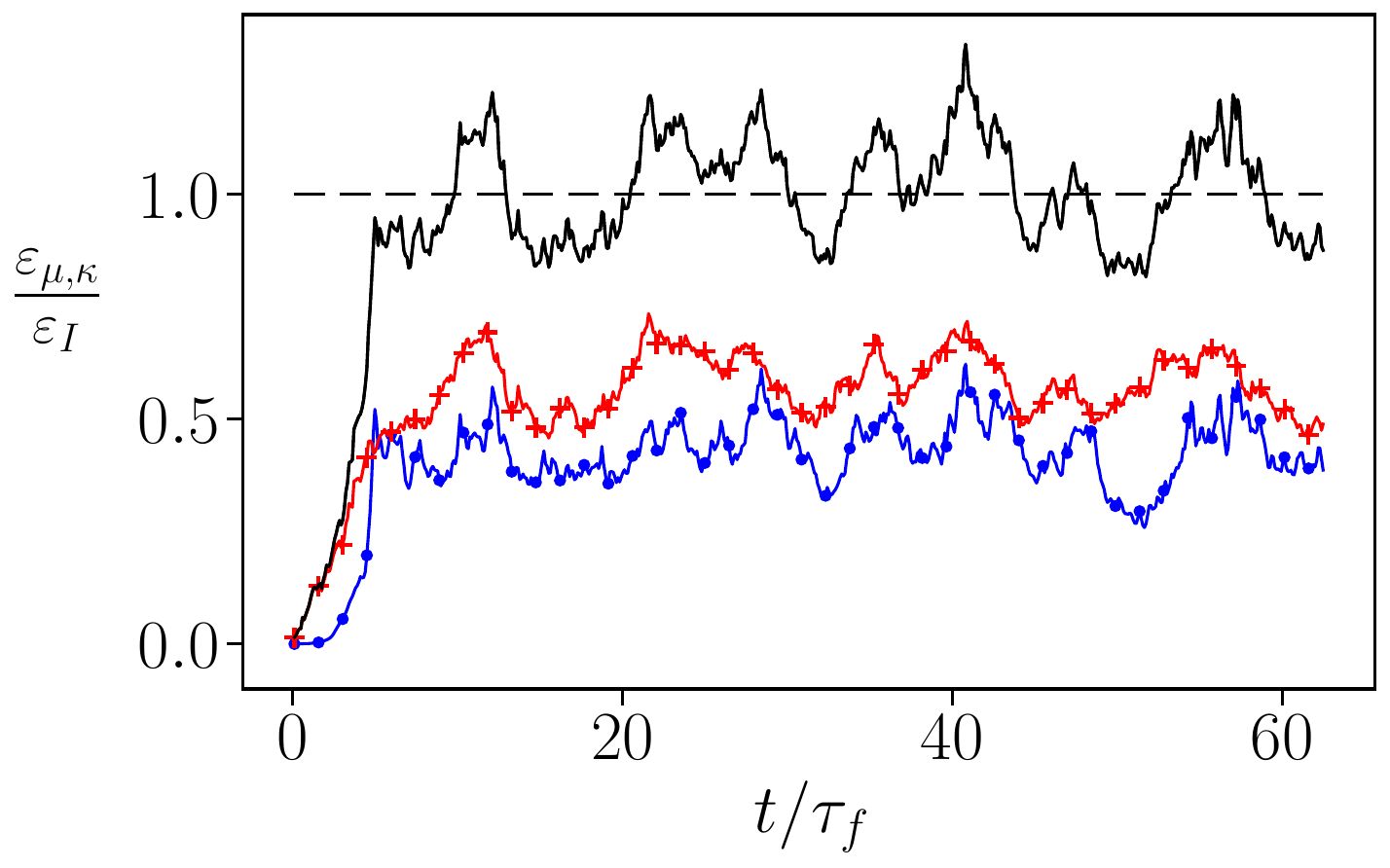}
\caption{Time evolution of the dissipative terms in the $P$-budget equation. The crossed line shows $\vare_\mu/\vare_I$, the bullet line shows $\vare_\kappa/\vare_I$ while the full line is the sum of the two. \label{fig3}}
\end{figure}

A more refined indicator of the stationarity of the system is provided
by the balance of dissipation and injection of SPE (i.e., of scalar variance).
In the presence of both large-scale and small-scale dissipative terms, the balance reads
$dP/dt = \varepsilon_I - \varepsilon_\kappa - \varepsilon_\mu$,
where $\varepsilon_\kappa = \kappa \langle |{\bm \nabla} \theta|^2 \rangle$
and  $\varepsilon_\mu = \mu \langle|{\bm \nabla}^{-1} \theta|^2 \rangle$
are the small-scale and large-scale dissipation rates, respectively,
and $\varepsilon_I = \langle \theta f \rangle$ is the SPE input rate.
In Figure~\ref{fig3} we show the temporal evolution of the dissipative terms
$\varepsilon_\kappa$ and $\varepsilon_\mu$.
We observe large fluctuations of the total dissipation $\varepsilon_\kappa+\varepsilon_\mu$,
which confirms that the stationarity condition $\varepsilon_\kappa+\varepsilon_\mu = \varepsilon_I$
is realized only in a statistical sense. 
We notice that a large fraction of the SPE injected by the forcing is immediately removed
at the forcing scale by hypo-friction terms, which prevents the development of the inverse cascade.
Therefore, the flux of the SPE in the direct cascade is only a fraction of the input
$\varepsilon_\kappa/\varepsilon_I \simeq 43\%$. 

The temporal evolution of $\varepsilon_\kappa$ and $\varepsilon_\mu$ can be
interpreted as proxies of the
small-scale and large-scale dynamics respectively. 
In particular, $\varepsilon_\mu$ shows an alternation of phases of growth, 
which correspond to the accumulation of SPE at large
scales, followed by phases of decrease.  The inversion occurs in correspondence
with intense dissipative events at small scales (the maxima of
$\varepsilon_\mu$). This shows that the system is never exactly at equilibrium.
The energy is gradually accumulated at large scales, until it is rapidly
discharged in the cascade causing a strong dissipative event at small scales,
then the process repeats. 

\begin{figure}[h]
\centering\includegraphics[width=0.7\linewidth]{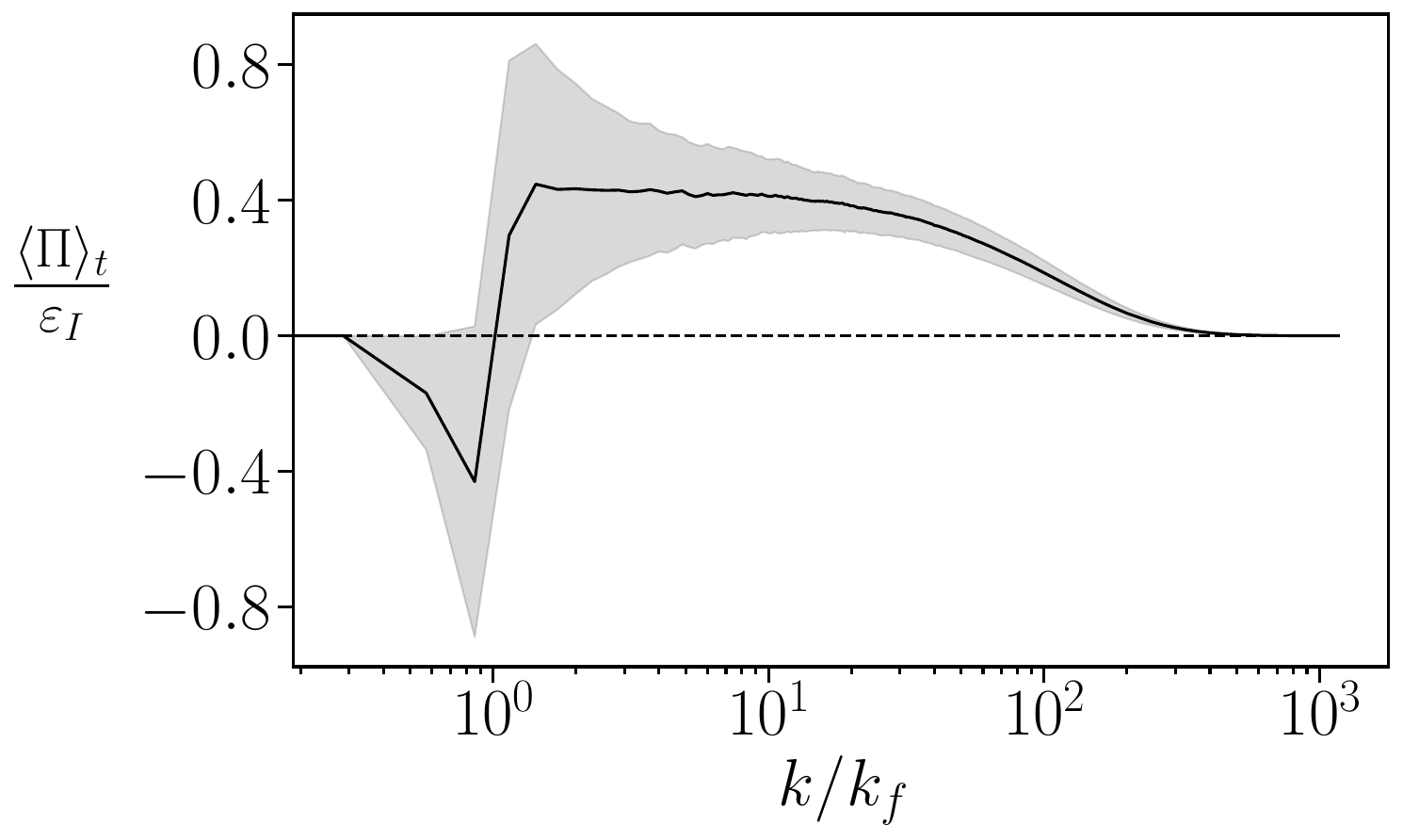}
\caption{Mean spectral flux $\mean{\Pi}_t$ across the wave number $k$,
  normalized by the input flux $\vare_I$ (black line).
  Shaded region shows its scale-by-scale $\sigma(k)$ fluctuation.}
\label{fig4}
\end{figure}

The strong instantaneous fluctuations of the cascade process are evident also in the spectral fluxes $\Pi(k,t)$. 
In Fig.~\ref{fig4} we show the time-averaged flux $\mean{\Pi(k)}_t$ across the circular shell of wavenumber $k$,
together with the region comprised within one standard deviation
$\sigma(k) = (\langle \Pi(k,t)^2\rangle_t - \langle \Pi(k,t)\rangle_t^2)^{1/2}$. 
The constancy of $\mean{\Pi}_t(k)$ along a range of scales that compress more than one decade
is in agreement with the assumption of an inertial cascade.
However, the large area covered within one standard deviation reveals the presence of significant
out-of-equilibrium fluctuations of the instantaneous fluxes $\Pi(k,t)$. 

\begin{figure}[h]
\centering\includegraphics[width=0.8\linewidth]{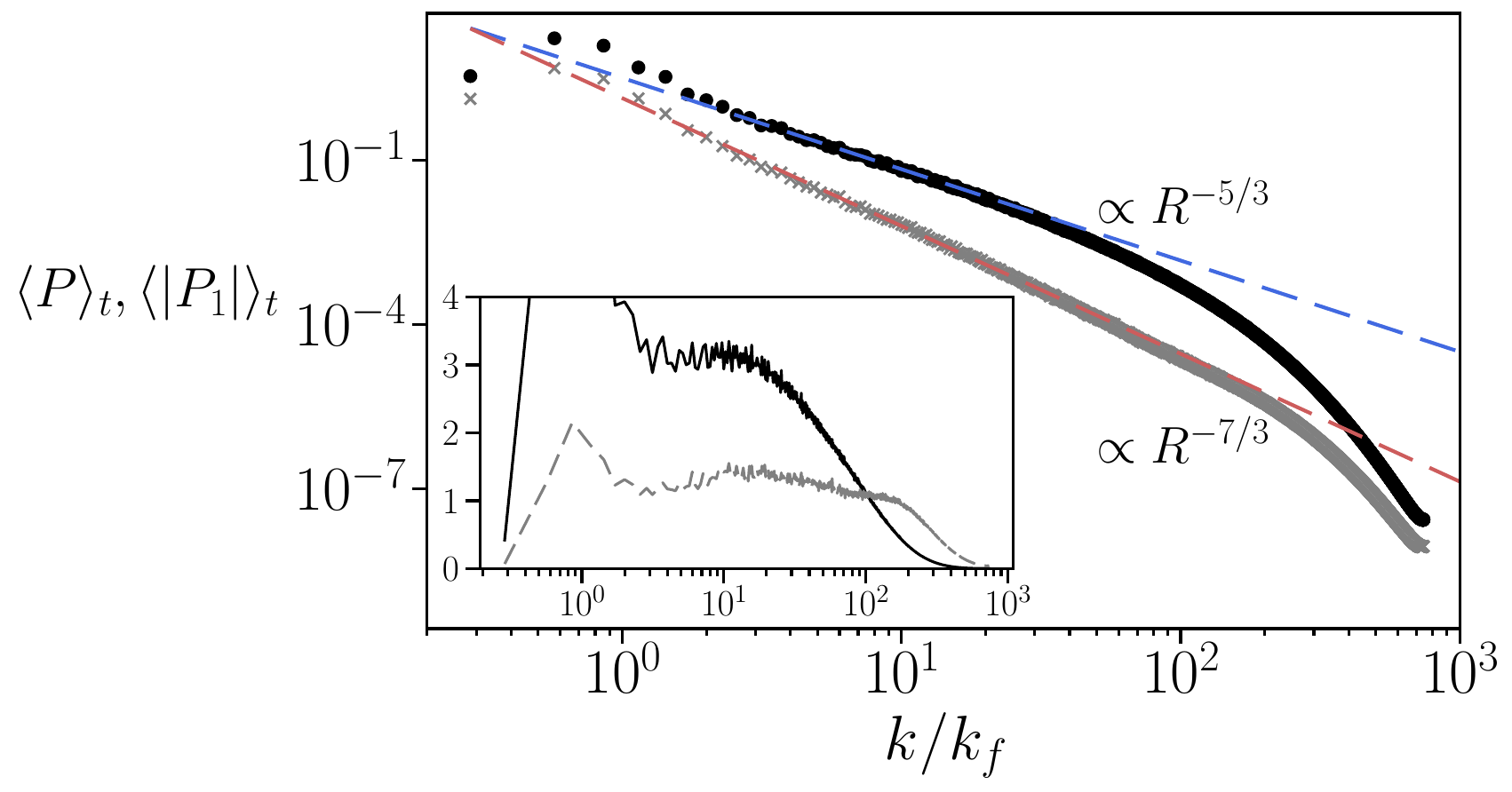}
\caption{Equilibrium spectrum (bullets) and non-equilibrium correction (marks) are shown in the main figure while the inset shows compensated spectra $\mean{P_0}_t(k/k_f)^{5/3}$ (full lines) and $\mean{|P_1|}_t(k/k_f)^{7/3}$ (dashed lines) as functions of $k/k_f$.}
\label{fig5}
\end{figure}

These fluctuations cause the appearance of spectral corrections
to the dimensional prediction for the equilibrium spectrum.
Following the procedure described in Section~\ref{sec2}, 
we computed the equilibrium
and non-equilibrium spectra, $P_0(k,t)$ and $P_1(k,t)$.  
In Fig.~\ref{fig5}, we show the time average
of the equilibrium spectrum $\mean{P_0(k,t)}_t$
and the time average of the absolute value of the non-equilibrium correction
$\mean{|P_1(k,t)|}_t$.
We remind that $\mean{P_1(k,t)}_t =0$,
therefore $\mean{P_0(k,t)}_t = \mean{P(k,t)}_t$.
For the mean equilibrium spectrum, we observe a good agreement with the dimensional prediction $\mean{P_0(k,t)}_t = C \mean{\vare_\kappa}_t^{2/3} k^{-5/3}$
with the dimensionless constant $C \simeq 5.14$.
We remark that previous work has found small corrections of the dimensional exponent \cite{celani2004active}.
In a set of preliminary simulations performed at smaller $Re_\kappa$ we observed similar corrections,
but the compensated plot in Fig.~\ref{fig5} shows that the numerical results at large $Re_\kappa$ 
are very close to the predicted scaling.
The spectral correction $P_1(k,t)$ is subdominant at all wavenumbers with
respect to the equilibrium spectrum, confirming a posteriori the validity of the expansion used in Sect.~\ref{sec2}.  Its spectral slope is in excellent agreement with $k^{-7/3}$ as predicted by Eq.~\eqref{eq12}. 

\begin{figure}[h]
    \includegraphics[width=0.7\linewidth]{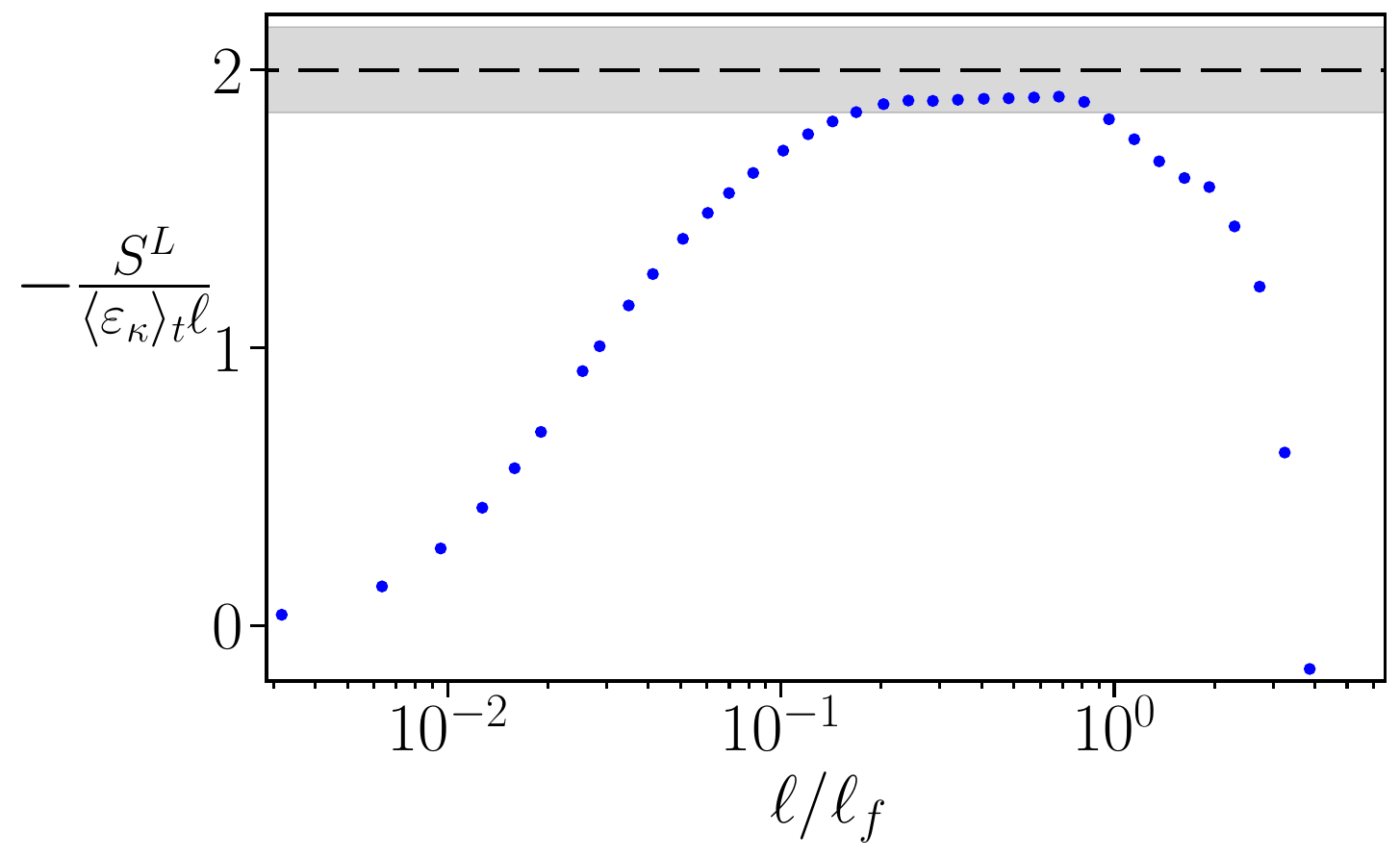}
\caption{Compensated third-order combined structure-function. The shaded area shows compatibility within $\pm 1$ standard deviation of $\vare_\kappa(t)$. \label{fig6}}
\end{figure}

Reinforcing our dimensional results and all previous derivations, we show in Fig.~\ref{fig6} the mixed longitudinal structure function $S^{L}(\ell)$ normalized following \eqref{eq6}. A clear plateau forms along roughly one decade with a value of about $5\%$ lower than expected. However, it is still compatible considering the effect of large temporal fluctuations of the small-scale dissipation $\mean{\vare_\kappa}_t$. 

\section{Conclusions}\label{conclusions}

We studied the effects of instantaneous out-of-equilibrium fluctuations
of the flux of turbulent cascades in the generalized model
of 2D transport equations known as $\alpha$-turbulence. 
Using a multiple scale approach, we derived a prediction,
valid for $0 < \alpha <2$, for the subleading correction to the energy spectrum
which originates from these fluctuations.
On the basis of these results, we propose an efficient method
for separating the equilibrium and non-equilibrium parts
in the instantaneous energy spectra. 

By means of high-resolution numerical simulations, we tested the predictions 
in the case of the Surface Quasi Geostrophic model, corresponding to $\alpha=1$.
Our results confirm the presence of large non-equilibrium temporal fluctuations
in the energy balance, accompanied by intense fluctuations of the spectral flux.
Both the time-averaged spectrum of the surface potential energy (i.e., of the scalar variance)
and the non-equilibrium subleading spectral correction
are found to be in agreement with the dimensional predictions $k^{-5/3}$ and $k^{-7/3}$, respectively. 
Moreover, we confirm the validity of the generalized Karman-Howarth-Monin equation for the mixed,
third-order structure function involving the temperature and longitudinal velocity differences. 

We note that, although the theoretical prediction for the spectral correction
is singular in the 2D Navier-Stokes case ($\alpha=2$), 
it is self consistent for all the values $0 < \alpha < 2$.
It would be interesting to test the general validity of this prediction
for other values of $\alpha$ in future numerical studies
of the $\alpha$-turbulence model. 

\section*{Acknowledgment}

We thank J. Bec, N. Valade for enlightening discussions on the SQG system.
S.M. thanks W. Bos for useful discussions on the non-equilibrium corrections. 
Supported by Italian Research Center on High Performance Computing Big Data and
Quantum Computing (ICSC), project funded by European Union - NextGenerationEU -
and National Recovery and Resilience Plan (NRRP) - Mission 4 Component 2 within
the activities of Spoke 3 (Astrophysics and Cosmos Observations).
We acknowledge HPC CINECA for computing resources within the INFN-CINECA Grant 
INFN24-FieldTurb.

\bibliography{biblio.bib}

\begin{thebibliography}{22}%
\makeatletter
\providecommand \@ifxundefined [1]{%
 \@ifx{#1\undefined}
}%
\providecommand \@ifnum [1]{%
 \ifnum #1\expandafter \@firstoftwo
 \else \expandafter \@secondoftwo
 \fi
}%
\providecommand \@ifx [1]{%
 \ifx #1\expandafter \@firstoftwo
 \else \expandafter \@secondoftwo
 \fi
}%
\providecommand \natexlab [1]{#1}%
\providecommand \enquote  [1]{``#1''}%
\providecommand \bibnamefont  [1]{#1}%
\providecommand \bibfnamefont [1]{#1}%
\providecommand \citenamefont [1]{#1}%
\providecommand \href@noop [0]{\@secondoftwo}%
\providecommand \href [0]{\begingroup \@sanitize@url \@href}%
\providecommand \@href[1]{\@@startlink{#1}\@@href}%
\providecommand \@@href[1]{\endgroup#1\@@endlink}%
\providecommand \@sanitize@url [0]{\catcode `\\12\catcode `\$12\catcode
  `\&12\catcode `\#12\catcode `\^12\catcode `\_12\catcode `\%12\relax}%
\providecommand \@@startlink[1]{}%
\providecommand \@@endlink[0]{}%
\providecommand \url  [0]{\begingroup\@sanitize@url \@url }%
\providecommand \@url [1]{\endgroup\@href {#1}{\urlprefix }}%
\providecommand \urlprefix  [0]{URL }%
\providecommand \Eprint [0]{\href }%
\providecommand \doibase [0]{https://doi.org/}%
\providecommand \selectlanguage [0]{\@gobble}%
\providecommand \bibinfo  [0]{\@secondoftwo}%
\providecommand \bibfield  [0]{\@secondoftwo}%
\providecommand \translation [1]{[#1]}%
\providecommand \BibitemOpen [0]{}%
\providecommand \bibitemStop [0]{}%
\providecommand \bibitemNoStop [0]{.\EOS\space}%
\providecommand \EOS [0]{\spacefactor3000\relax}%
\providecommand \BibitemShut  [1]{\csname bibitem#1\endcsname}%
\let\auto@bib@innerbib\@empty
\bibitem [{\citenamefont {Blumen}(1978)}]{blumen1978uniform}%
  \BibitemOpen
  \bibfield  {author} {\bibinfo {author} {\bibfnamefont {W.}~\bibnamefont
  {Blumen}},\ }\bibfield  {title} {\bibinfo {title} {{Uniform potential
  vorticity flow: Part I. Theory of wave interactions and two-dimensional
  turbulence}},\ }\href@noop {} {\bibfield  {journal} {\bibinfo  {journal} {J.
  Atmos. Sciences}\ }\textbf {\bibinfo {volume} {35}},\ \bibinfo {pages} {774}
  (\bibinfo {year} {1978})}\BibitemShut {NoStop}%
\bibitem [{\citenamefont {Salmon}(1998)}]{salmon1998lectures}%
  \BibitemOpen
  \bibfield  {author} {\bibinfo {author} {\bibfnamefont {R.}~\bibnamefont
  {Salmon}},\ }\href@noop {} {\emph {\bibinfo {title} {Lectures on geophysical
  fluid dynamics}}}\ (\bibinfo  {publisher} {Oxford University Press, USA},\
  \bibinfo {year} {1998})\BibitemShut {NoStop}%
\bibitem [{\citenamefont {Juckes}(1994)}]{juckes1994quasigeostrophic}%
  \BibitemOpen
  \bibfield  {author} {\bibinfo {author} {\bibfnamefont {M.}~\bibnamefont
  {Juckes}},\ }\bibfield  {title} {\bibinfo {title} {Quasigeostrophic dynamics
  of the tropopause},\ }\href@noop {} {\bibfield  {journal} {\bibinfo
  {journal} {J. Atmos. Sciences}\ }\textbf {\bibinfo {volume} {51}},\ \bibinfo
  {pages} {2756} (\bibinfo {year} {1994})}\BibitemShut {NoStop}%
\bibitem [{\citenamefont {Lapeyre}\ and\ \citenamefont
  {Klein}(2006)}]{lapeyre2006dynamics}%
  \BibitemOpen
  \bibfield  {author} {\bibinfo {author} {\bibfnamefont {G.}~\bibnamefont
  {Lapeyre}}\ and\ \bibinfo {author} {\bibfnamefont {P.}~\bibnamefont
  {Klein}},\ }\bibfield  {title} {\bibinfo {title} {Dynamics of the upper
  oceanic layers in terms of surface quasigeostrophy theory},\ }\href@noop {}
  {\bibfield  {journal} {\bibinfo  {journal} {J. Phys. Ocean.}\ }\textbf
  {\bibinfo {volume} {36}},\ \bibinfo {pages} {165} (\bibinfo {year}
  {2006})}\BibitemShut {NoStop}%
\bibitem [{\citenamefont {Siegelman}\ \emph {et~al.}(2022)\citenamefont
  {Siegelman}, \citenamefont {Klein}, \citenamefont {Ingersoll}, \citenamefont
  {Ewald}, \citenamefont {Young}, \citenamefont {Bracco}, \citenamefont {Mura},
  \citenamefont {Adriani}, \citenamefont {Grassi}, \citenamefont {Plainaki}
  \emph {et~al.}}]{siegelman2022moist}%
  \BibitemOpen
  \bibfield  {author} {\bibinfo {author} {\bibfnamefont {L.}~\bibnamefont
  {Siegelman}}, \bibinfo {author} {\bibfnamefont {P.}~\bibnamefont {Klein}},
  \bibinfo {author} {\bibfnamefont {A.~P.}\ \bibnamefont {Ingersoll}}, \bibinfo
  {author} {\bibfnamefont {S.~P.}\ \bibnamefont {Ewald}}, \bibinfo {author}
  {\bibfnamefont {W.~R.}\ \bibnamefont {Young}}, \bibinfo {author}
  {\bibfnamefont {A.}~\bibnamefont {Bracco}}, \bibinfo {author} {\bibfnamefont
  {A.}~\bibnamefont {Mura}}, \bibinfo {author} {\bibfnamefont {A.}~\bibnamefont
  {Adriani}}, \bibinfo {author} {\bibfnamefont {D.}~\bibnamefont {Grassi}},
  \bibinfo {author} {\bibfnamefont {C.}~\bibnamefont {Plainaki}}, \emph
  {et~al.},\ }\bibfield  {title} {\bibinfo {title} {{Moist convection drives an
  upscale energy transfer at Jovian high latitudes}},\ }\href@noop {}
  {\bibfield  {journal} {\bibinfo  {journal} {Nature Physics}\ }\textbf
  {\bibinfo {volume} {18}},\ \bibinfo {pages} {357} (\bibinfo {year}
  {2022})}\BibitemShut {NoStop}%
\bibitem [{\citenamefont {Pierrehumbert}\ \emph {et~al.}(1994)\citenamefont
  {Pierrehumbert}, \citenamefont {Held},\ and\ \citenamefont
  {Swanson}}]{pierrehumbert1994spectra}%
  \BibitemOpen
  \bibfield  {author} {\bibinfo {author} {\bibfnamefont {R.~T.}\ \bibnamefont
  {Pierrehumbert}}, \bibinfo {author} {\bibfnamefont {I.~M.}\ \bibnamefont
  {Held}},\ and\ \bibinfo {author} {\bibfnamefont {K.~L.}\ \bibnamefont
  {Swanson}},\ }\bibfield  {title} {\bibinfo {title} {Spectra of local and
  nonlocal two-dimensional turbulence},\ }\href@noop {} {\bibfield  {journal}
  {\bibinfo  {journal} {Chaos, Solitons \& Fractals}\ }\textbf {\bibinfo
  {volume} {4}},\ \bibinfo {pages} {1111} (\bibinfo {year} {1994})}\BibitemShut
  {NoStop}%
\bibitem [{\citenamefont {Held}\ \emph {et~al.}(1995)\citenamefont {Held},
  \citenamefont {Pierrehumbert}, \citenamefont {Garner},\ and\ \citenamefont
  {Swanson}}]{held1995surface}%
  \BibitemOpen
  \bibfield  {author} {\bibinfo {author} {\bibfnamefont {I.~M.}\ \bibnamefont
  {Held}}, \bibinfo {author} {\bibfnamefont {R.~T.}\ \bibnamefont
  {Pierrehumbert}}, \bibinfo {author} {\bibfnamefont {S.~T.}\ \bibnamefont
  {Garner}},\ and\ \bibinfo {author} {\bibfnamefont {K.~L.}\ \bibnamefont
  {Swanson}},\ }\bibfield  {title} {\bibinfo {title} {Surface quasi-geostrophic
  dynamics},\ }\href@noop {} {\bibfield  {journal} {\bibinfo  {journal} {J.
  Fluid Mech.}\ }\textbf {\bibinfo {volume} {282}},\ \bibinfo {pages} {1}
  (\bibinfo {year} {1995})}\BibitemShut {NoStop}%
\bibitem [{\citenamefont {Celani}\ \emph {et~al.}(2004)\citenamefont {Celani},
  \citenamefont {Cencini}, \citenamefont {Mazzino},\ and\ \citenamefont
  {Vergassola}}]{celani2004active}%
  \BibitemOpen
  \bibfield  {author} {\bibinfo {author} {\bibfnamefont {A.}~\bibnamefont
  {Celani}}, \bibinfo {author} {\bibfnamefont {M.}~\bibnamefont {Cencini}},
  \bibinfo {author} {\bibfnamefont {A.}~\bibnamefont {Mazzino}},\ and\ \bibinfo
  {author} {\bibfnamefont {M.}~\bibnamefont {Vergassola}},\ }\bibfield  {title}
  {\bibinfo {title} {Active and passive fields face to face},\ }\href@noop {}
  {\bibfield  {journal} {\bibinfo  {journal} {New J. Phys.}\ }\textbf {\bibinfo
  {volume} {6}},\ \bibinfo {pages} {72} (\bibinfo {year} {2004})}\BibitemShut
  {NoStop}%
\bibitem [{\citenamefont {Lapeyre}(2017)}]{lapeyre2017surface}%
  \BibitemOpen
  \bibfield  {author} {\bibinfo {author} {\bibfnamefont {G.}~\bibnamefont
  {Lapeyre}},\ }\bibfield  {title} {\bibinfo {title} {Surface
  quasi-geostrophy},\ }\href@noop {} {\bibfield  {journal} {\bibinfo  {journal}
  {Fluids}\ }\textbf {\bibinfo {volume} {2}},\ \bibinfo {pages} {7} (\bibinfo
  {year} {2017})}\BibitemShut {NoStop}%
\bibitem [{\citenamefont {Foussard}\ \emph {et~al.}(2017)\citenamefont
  {Foussard}, \citenamefont {Berti}, \citenamefont {Perrot},\ and\
  \citenamefont {Lapeyre}}]{foussard2017relative}%
  \BibitemOpen
  \bibfield  {author} {\bibinfo {author} {\bibfnamefont {A.}~\bibnamefont
  {Foussard}}, \bibinfo {author} {\bibfnamefont {S.}~\bibnamefont {Berti}},
  \bibinfo {author} {\bibfnamefont {X.}~\bibnamefont {Perrot}},\ and\ \bibinfo
  {author} {\bibfnamefont {G.}~\bibnamefont {Lapeyre}},\ }\bibfield  {title}
  {\bibinfo {title} {Relative dispersion in generalized two-dimensional
  turbulence},\ }\href@noop {} {\bibfield  {journal} {\bibinfo  {journal} {J.
  Fluid Mech.}\ }\textbf {\bibinfo {volume} {821}},\ \bibinfo {pages} {358}
  (\bibinfo {year} {2017})}\BibitemShut {NoStop}%
\bibitem [{\citenamefont {Constantin}\ \emph {et~al.}(1994)\citenamefont
  {Constantin}, \citenamefont {Majda},\ and\ \citenamefont
  {Tabak}}]{constantin1994formation}%
  \BibitemOpen
  \bibfield  {author} {\bibinfo {author} {\bibfnamefont {P.}~\bibnamefont
  {Constantin}}, \bibinfo {author} {\bibfnamefont {A.~J.}\ \bibnamefont
  {Majda}},\ and\ \bibinfo {author} {\bibfnamefont {E.}~\bibnamefont {Tabak}},\
  }\bibfield  {title} {\bibinfo {title} {Formation of strong fronts in the 2-d
  quasigeostrophic thermal active scalar},\ }\href@noop {} {\bibfield
  {journal} {\bibinfo  {journal} {Nonlinearity}\ }\textbf {\bibinfo {volume}
  {7}},\ \bibinfo {pages} {1495} (\bibinfo {year} {1994})}\BibitemShut
  {NoStop}%
\bibitem [{\citenamefont {Constantin}\ and\ \citenamefont
  {Wu}(1999)}]{constantin1999behavior}%
  \BibitemOpen
  \bibfield  {author} {\bibinfo {author} {\bibfnamefont {P.}~\bibnamefont
  {Constantin}}\ and\ \bibinfo {author} {\bibfnamefont {J.}~\bibnamefont
  {Wu}},\ }\bibfield  {title} {\bibinfo {title} {Behavior of solutions of 2d
  quasi-geostrophic equations},\ }\href@noop {} {\bibfield  {journal} {\bibinfo
   {journal} {SIAM J. Math. Analys.}\ }\textbf {\bibinfo {volume} {30}},\
  \bibinfo {pages} {937} (\bibinfo {year} {1999})}\BibitemShut {NoStop}%
\bibitem [{\citenamefont {Valade}\ \emph {et~al.}(2024)\citenamefont {Valade},
  \citenamefont {Thalabard},\ and\ \citenamefont {Bec}}]{valade2024anomalous}%
  \BibitemOpen
  \bibfield  {author} {\bibinfo {author} {\bibfnamefont {N.}~\bibnamefont
  {Valade}}, \bibinfo {author} {\bibfnamefont {S.}~\bibnamefont {Thalabard}},\
  and\ \bibinfo {author} {\bibfnamefont {J.}~\bibnamefont {Bec}},\ }\bibfield
  {title} {\bibinfo {title} {Anomalous dissipation and spontaneous
  stochasticity in deterministic surface quasi-geostrophic flow},\ }in\
  \href@noop {} {\emph {\bibinfo {booktitle} {Annales Henri Poincar{\'e}}}},\
  Vol.~\bibinfo {volume} {25}\ (\bibinfo {organization} {Springer},\ \bibinfo
  {year} {2024})\ pp.\ \bibinfo {pages} {1261--1283}\BibitemShut {NoStop}%
\bibitem [{\citenamefont {Yoshizawa}(1994)}]{yoshizawa1994nonequilibrium}%
  \BibitemOpen
  \bibfield  {author} {\bibinfo {author} {\bibfnamefont {A.}~\bibnamefont
  {Yoshizawa}},\ }\bibfield  {title} {\bibinfo {title} {Nonequilibrium effect
  of the turbulent-energy-production process on the inertial-range energy
  spectrum},\ }\href@noop {} {\bibfield  {journal} {\bibinfo  {journal} {Phys.
  Rev. E}\ }\textbf {\bibinfo {volume} {49}},\ \bibinfo {pages} {4065}
  (\bibinfo {year} {1994})}\BibitemShut {NoStop}%
\bibitem [{\citenamefont {Woodruff}\ and\ \citenamefont
  {Rubinstein}(2006)}]{woodruff2006multiple}%
  \BibitemOpen
  \bibfield  {author} {\bibinfo {author} {\bibfnamefont {S.~L.}\ \bibnamefont
  {Woodruff}}\ and\ \bibinfo {author} {\bibfnamefont {R.}~\bibnamefont
  {Rubinstein}},\ }\bibfield  {title} {\bibinfo {title} {Multiple-scale
  perturbation analysis of slowly evolving turbulence},\ }\href@noop {}
  {\bibfield  {journal} {\bibinfo  {journal} {J. Fluid Mech.}\ }\textbf
  {\bibinfo {volume} {565}},\ \bibinfo {pages} {95} (\bibinfo {year}
  {2006})}\BibitemShut {NoStop}%
\bibitem [{\citenamefont {Berti}\ \emph {et~al.}(2023)\citenamefont {Berti},
  \citenamefont {Boffetta},\ and\ \citenamefont {Musacchio}}]{berti2023mean}%
  \BibitemOpen
  \bibfield  {author} {\bibinfo {author} {\bibfnamefont {S.}~\bibnamefont
  {Berti}}, \bibinfo {author} {\bibfnamefont {G.}~\bibnamefont {Boffetta}},\
  and\ \bibinfo {author} {\bibfnamefont {S.}~\bibnamefont {Musacchio}},\
  }\bibfield  {title} {\bibinfo {title} {Mean flow and fluctuations in the
  three-dimensional turbulent cellular flow},\ }\href@noop {} {\bibfield
  {journal} {\bibinfo  {journal} {Phys. Rev. Fluids}\ }\textbf {\bibinfo
  {volume} {8}},\ \bibinfo {pages} {054601} (\bibinfo {year}
  {2023})}\BibitemShut {NoStop}%
\bibitem [{\citenamefont {Frisch}(1995)}]{frisch1995turbulence}%
  \BibitemOpen
  \bibfield  {author} {\bibinfo {author} {\bibfnamefont {U.}~\bibnamefont
  {Frisch}},\ }\href@noop {} {\emph {\bibinfo {title} {{Turbulence: the legacy
  of A.N. Kolmogorov}}}}\ (\bibinfo  {publisher} {Cambridge University Press},\
  \bibinfo {year} {1995})\BibitemShut {NoStop}%
\bibitem [{\citenamefont {Bos}\ and\ \citenamefont
  {Rubinstein}(2017)}]{bos2017dissipation}%
  \BibitemOpen
  \bibfield  {author} {\bibinfo {author} {\bibfnamefont {W.~J.}\ \bibnamefont
  {Bos}}\ and\ \bibinfo {author} {\bibfnamefont {R.}~\bibnamefont
  {Rubinstein}},\ }\bibfield  {title} {\bibinfo {title} {Dissipation in
  unsteady turbulence},\ }\href@noop {} {\bibfield  {journal} {\bibinfo
  {journal} {Phys. Rev. Fluids}\ }\textbf {\bibinfo {volume} {2}},\ \bibinfo
  {pages} {022601(R)} (\bibinfo {year} {2017})}\BibitemShut {NoStop}%
\bibitem [{\citenamefont {Batchelor}(1953)}]{batchelor1953theory}%
  \BibitemOpen
  \bibfield  {author} {\bibinfo {author} {\bibfnamefont {G.~K.}\ \bibnamefont
  {Batchelor}},\ }\href@noop {} {\emph {\bibinfo {title} {The theory of
  homogeneous turbulence}}}\ (\bibinfo  {publisher} {Cambridge university
  press},\ \bibinfo {year} {1953})\BibitemShut {NoStop}%
\bibitem [{\citenamefont {Kovasznay}(1948)}]{kovasznay1948spectrum}%
  \BibitemOpen
  \bibfield  {author} {\bibinfo {author} {\bibfnamefont {L.~S.}\ \bibnamefont
  {Kovasznay}},\ }\bibfield  {title} {\bibinfo {title} {Spectrum of locally
  isotropic turbulence},\ }\href@noop {} {\bibfield  {journal} {\bibinfo
  {journal} {J. Aeron. Sciences}\ }\textbf {\bibinfo {volume} {15}},\ \bibinfo
  {pages} {745} (\bibinfo {year} {1948})}\BibitemShut {NoStop}%
\bibitem [{\citenamefont {Boyd}(2001)}]{boyd2001chebyshev}%
  \BibitemOpen
  \bibfield  {author} {\bibinfo {author} {\bibfnamefont {J.~P.}\ \bibnamefont
  {Boyd}},\ }\href@noop {} {\emph {\bibinfo {title} {{Chebyshev and Fourier
  spectral methods}}}}\ (\bibinfo  {publisher} {Courier Corporation},\ \bibinfo
  {year} {2001})\BibitemShut {NoStop}%
\bibitem [{\citenamefont {Valad\~ao}\ \emph {et~al.}(2024)\citenamefont
  {Valad\~ao}, \citenamefont {Boffetta}, \citenamefont {Crialesi-Esposito},
  \citenamefont {De~Lillo},\ and\ \citenamefont
  {Musacchio}}]{valadao2024spectrum}%
  \BibitemOpen
  \bibfield  {author} {\bibinfo {author} {\bibfnamefont {V.~J.}\ \bibnamefont
  {Valad\~ao}}, \bibinfo {author} {\bibfnamefont {G.}~\bibnamefont {Boffetta}},
  \bibinfo {author} {\bibfnamefont {M.}~\bibnamefont {Crialesi-Esposito}},
  \bibinfo {author} {\bibfnamefont {F.}~\bibnamefont {De~Lillo}},\ and\
  \bibinfo {author} {\bibfnamefont {S.}~\bibnamefont {Musacchio}},\ }\bibfield
  {title} {\bibinfo {title} {{Spectrum correction on Ekman-Navier-Stokes
  equation in two-dimensions}},\ }\href@noop {} {\bibfield  {journal} {\bibinfo
   {journal} {in preparation}\ } (\bibinfo {year} {2024})}\BibitemShut
  {NoStop}%
\end{thebibliography}%

\clearpage

\appendix*

\section{Appendix A}
\label{app:a}
It is possible to derive an exact relation for the flux of the transported
field $\theta(x,t)$ in the general model (\ref{eq1}) in stationary conditions
and under the assumption of homogeneity and isotropy.
For simplicity, we introduce shortened notations
like $\theta'\equiv\theta(\bm x',t)$, where $\bm{x}'=\bm{x}+\bm{\ell}$
and we remind that, as a consequence of homogeneity, we have
$\nabla_i\mean{\lr{.}}=-\nabla'_i\mean{\lr{.}}=-\nabla_{\ell_i}\mean{\lr{.}}$.
We start from the time evolution of the two-point correlation
correlations, 
\begin{equation}
\partial_t\mean{\theta\theta'}=\mean{\theta'\partial_t\theta}+\mean{\theta\partial_t\theta'}
\label{eqa1}
\end{equation}
which, by the use of Eq.~\eqref{eq1} reads,
\begin{align}
\partial_t\mean{\theta'\theta}=-&\lr{\mean{\theta'\nabla_i\lr{v_i\theta}}+\mean{\theta\nabla'_i\lr{v'_i\theta'}}}+ \nonumber\\
+&\kappa\lr{\mean{\theta'\nabla^2\theta}+\mean{\theta\nabla^2\theta'}}+ \nonumber\\
-&\mu\lr{\mean{\theta'\nabla^{-2}\theta}+\mean{\theta\nabla^{-2}\theta'}}+ \nonumber\\
+&\mean{\theta'f}+\mean{\theta f'}
\label{eqa2}
\end{align}
where we have included the hypo-friction term discussed in 
Section~\ref{sec4} and the sum over repeated indices is implied. The forcing term defines the injection rate
of scalar variance defined as $\Pi(\ell) \equiv \mean{\theta f'}$.
Assuming stationarity conditions and neglecting the contribution of 
dissipative terms for inertial range scales $\ell_\kappa\ll\ell\ll\ell_f$, one obtains
\begin{equation}
\Pi(\ell)=\frac{1}{2}\nabla_{\ell_i}\Big(\mean{\theta v'_i\theta'}-\mean{\theta'v_i\theta}\Big)
\label{eqa3}
\end{equation}
The RHS of this expression can be rewritten in terms of Galilean 
invariant observables, i.e. spatial increments such as 
$\delta_\ell\theta\equiv\theta'-\theta$.
Indeed we can write 
\begin{align}
\mean{(\delta\theta)^2\delta v_i}=&\mean{(\theta^2+\theta'^2-2\theta\theta')(v'_i-v_i)}= \nonumber\\
=&\mean{\theta^2v_i'}-\mean{\theta'^2v_i}+ \nonumber\\
+&2\mean{\theta'\theta v_i}-2\mean{\theta\theta'v_i'}
\label{eqa4}
\end{align}
and by taking the divergence over the $\ell$ variable in the above
equation, the first two terms vanish due to incompressibility, and the
remaining terms are written as
\begin{equation}
\nabla_{\ell_i}\mean{(\delta\theta)^2\delta v_i}=2\nabla_{\ell_i}\Big(\mean{\theta'\theta v_i}-\mean{\theta\theta'v_i'}\Big)
\label{eqa5}
\end{equation}
in which, by the use of Eq.~\eqref{eqa3}, one gets
\begin{equation}
\Pi(\ell)=-\frac{1}{4}\nabla_{\ell_i}\mean{(\delta\theta)^2\delta v_i} \ .\
\label{eqa6}
\end{equation}
Now, assuming isotropy, one has that the most general tensorial structure 
for $\mean{(\delta\theta)^2\delta v_i}$ is 
\begin{equation}
\mean{(\delta\theta)^2\delta v_i}=g(\ell)\frac{\ell_i}{\ell}
\label{eqa7}
\end{equation}
where $g(\ell)$ is a function of $\ell=|\bm\ell_i|$ who is directly identified
to the mixed longitudinal structure function
$S^L(\ell)=g(\ell)=\mean{(\delta\theta)^2\delta v_i}\ell_i/\ell$. 

Finally, assuming that over the range of scale in which dissipations can
be neglected the flux $\Pi(\ell)$ is transferred at a constant rate 
$\vare$ we have
\begin{equation}
\Pi(\ell)=-\frac{1}{4}\nabla_{\ell_i}\lr{S^L(\ell)\frac{\ell_i}{\ell}}\approx \vare
\label{eqa8}
\end{equation}
which can be written in the following form,
\begin{equation}
\frac{dS^L}{d\ell}+\frac{S^L}{\ell}=-4\vare
\label{eqa9}
\end{equation}
As a first-order differential equation, constrained by UV convergence, since
increments are Galilean invariant, or mathematically 
$S^L(\ell\rightarrow 0)=0$, the unique solution is given by
\begin{equation}
S^L(\ell)=-2\vare\ell \ .\
\label{eqa10}
\end{equation}
It is worth emphasizing that the derivation of equation \eqref{eqa10} is completely independent of the intrinsic relation between $\theta$ and $\psi$, being valid in the case of standard NS turbulence or in the generic case where $\hat\theta_k=f(k)\hat\psi_k$ comprising the $\alpha$-turbulence model and even more complicated transport equations.

\end{document}